\documentclass[pre,twocolumn,superscriptaddress,floatfix,showpacs,showkeys]{revtex4}

\usepackage[latin2]{inputenc}
\usepackage{amsmath}
\usepackage{amssymb}
\usepackage{epsfig}

\newcommand{\Fcrit}{F_\text{crit}}
\newcommand{\Fcont}{F_\text{cont}}

\begin{document}

\title{Unjamming due to local perturbations in granular packings
with and without gravity}

\date{\today}

\author{M.\ Reza Shaebani}
\email{shaebani@iasbs.ac.ir} 
\affiliation{Department of Theoretical Physics, Budapest University of
 Technology and Economics, H-1111 Budapest, Hungary}
 \affiliation{Institute for Advanced Studies in Basic Sciences,
 Zanjan 45195-1159, Iran}
\author{Tam\'as Unger}
\affiliation{Department of Theoretical Physics, Budapest University of
 Technology and Economics, H-1111 Budapest, Hungary}
\affiliation{HAS-BME Condensed Matter Research Group, Budapest 
University of Technology and Economics}
\author{J\'anos Kert\'esz}
\affiliation{Department of Theoretical Physics, Budapest University of
 Technology and Economics, H-1111 Budapest, Hungary}
\affiliation{HAS-BME Condensed Matter Research Group, Budapest 
University of Technology and Economics}

\begin{abstract}
We investigate the unjamming response of disordered packings of
frictional hard disks with the help of computer simulations.
First, we generate jammed configurations of the disks and then
force them to move again by local perturbations. We study the
spatial distribution of the stress and displacement response and
find long range effects of the perturbation in both cases. We
record the penetration depth of the displacements and the critical
force that is needed to make the system yield. These quantities
are tested in two types of systems: in ideal homogeneous packings
in zero gravity and in packings settled under gravity. The
penetration depth and the critical force are sensitive to the
interparticle friction coefficient. Qualitatively, the same
nonmonotonic friction dependence is found both with and without
gravity, however the location of the extrema are at different
friction values. We discuss the role of the connectivity of the
contact network and of the pressure gradient in the unjamming
response.

\end{abstract}



\maketitle

\section{Introduction}

Many systems including granular materials, foams and emulsions
can flow like fluids when a high external stress is applied but
jam into a solidlike state below a certain threshold of stress.
In a jammed state \cite{Liu98,Liu01,Makse04,Kolb04} the many-body
system is trapped in a metastable configuration far from
equilibrium where the constituent particles block each others
motion. For a typical jammed granular packing, where the thermal
fluctuations are negligible, only a sufficiently high external
stress can lead to an unjamming transition and cause 
rearrangements of the particles.

In this paper we study the unjamming response of dense disordered
granular media based on computer simulations. To achieve the
unjamming transition we perturb the material by generating a small
local deformation. These perturbations break the static structure
of the packings and induce motion of particles.

The response of granular media to local perturbations have been
studied widely both in experiments
\cite{Geng01,Kolb04,Serero01,Kolb06,Atman05} and in computer
simulations \cite{Atman05,Goldenberg05,Ostojic06,Goldenberg06}. The majority of
these studies apply small force overloads to study the stress
response inside the bulk of material. In these cases, the
displacements originate only from elastic deformation of the
particles; the system remains in the jammed state. Unjamming
induced by local perturbation has been also investigated
experimentally. Kolb and co-workers \cite{Kolb04,Kolb06} studied two-dimensional 
packings of disks under gravity and applied localized cyclic
perturbations to achieve real plastic rearrangements. Based on the
displacement field they showed that the unjamming response is long
ranged; the magnitude of the particle displacements decays as a
power law of the distance from the perturbation source. The
exponent of the decay varied between $0.6$ and $1.4$ depending on
the preparation of the system.

\begin{figure} [b]
\epsfig{figure=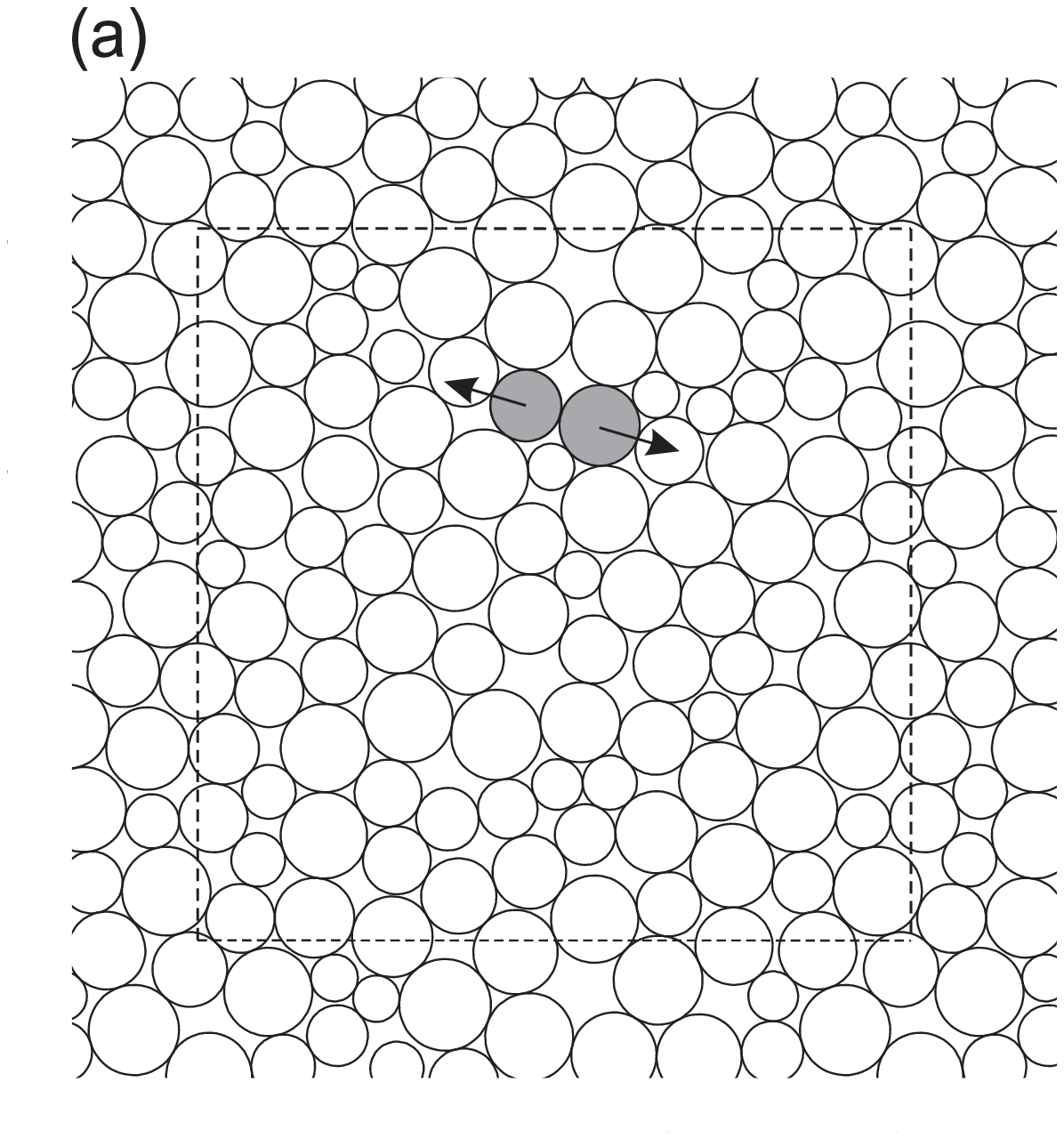,width=0.47\linewidth}
\epsfig{figure=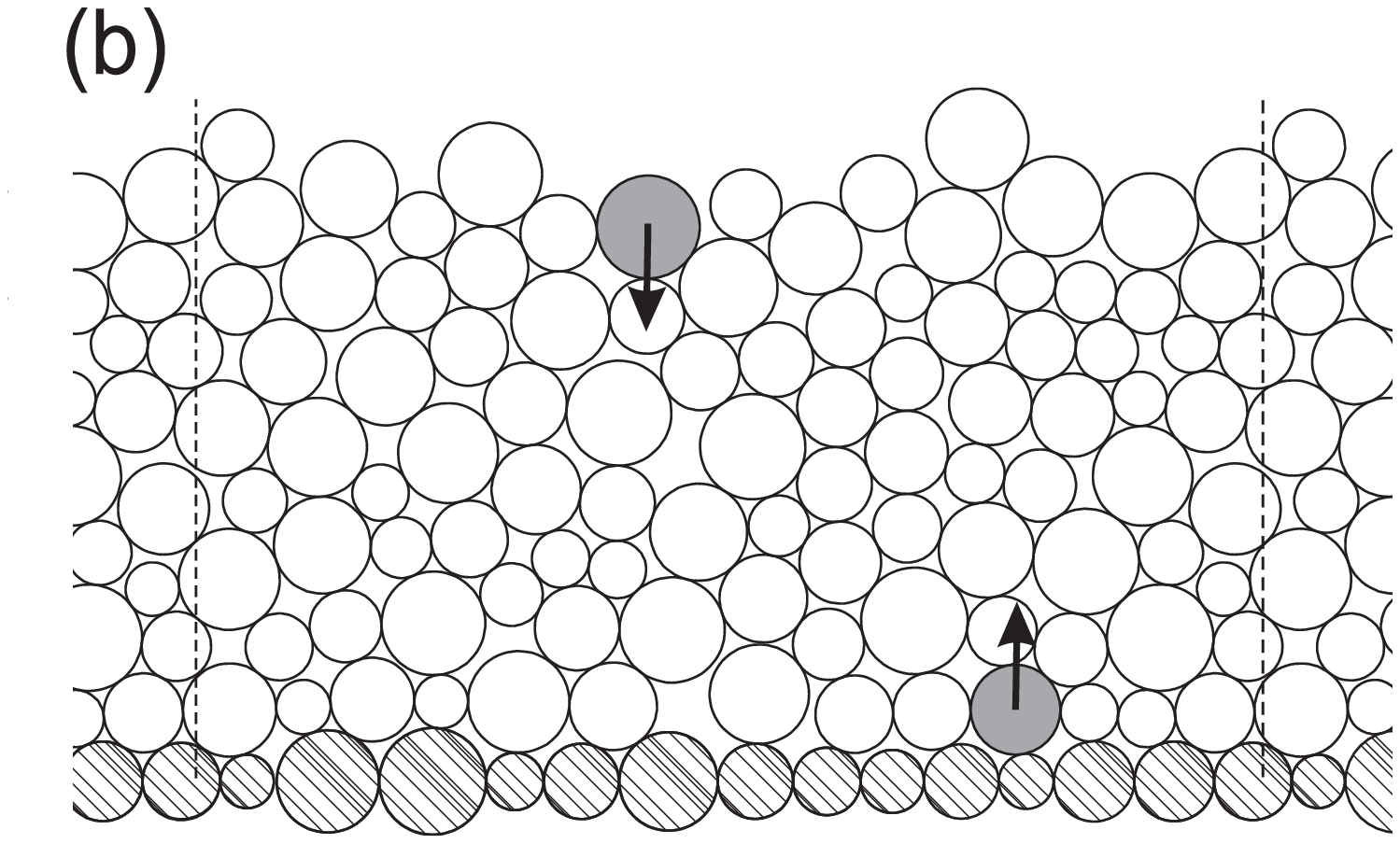,width=0.50\linewidth}
\caption{Schematic picture of the two types of granular packings
confined by an external pressure bath (a) and by gravity (b). The
dashed lines mark periodic boundaries. Some typical perturbations
are illustrated with gray particles and arrows showing the
direction of perturbation.} {\label{Fig-SchematicPicture}}
\end{figure}

Here we focus on the question of what role the interparticle friction
and gravity play in the unjamming transition. We study the
unjamming response with the help of the contact dynamics method 
\cite{Moreau94,Jean99,Brendel04} which handles the
particles as perfectly rigid bodies. Therefore elastic distortion of the
particles are excluded and the generated particle movements
correspond to real plastic rearrangements. We analyze the response
of the packing to local perturbations by considering the
individual particle displacements, the coarse-grained displacement
field, and, in addition, the resistance of the system against the
deformation. We show that the stability of the packing against
local perturbations depends strongly and in a nontrivial way on
the particle-particle friction coefficient. Our systems are 
two-dimensional disordered packings of disks. First, we test ideal
homogeneous packings prepared by an isotropic external pressure
and with fully periodic boundary conditions
\cite{Shaebani07,Shaebani08jcp}
[Fig.~\ref{Fig-SchematicPicture}(a)]. This part is the full
exposition and expansion of the results presented and considered
from another point of view in \cite{Shaebani07}. Then we study
more realistic packings that are settled under gravity
[Fig.~\ref{Fig-SchematicPicture}(b)]. The packings are perturbed
by moving two adjacent particles apart in the former case and by
shifting one particle vertically in the latter case.

This work is organized in the following manner:
Section~\ref{SimulationMethod-Section} contains the description of
the simulation method. In Sec.~\ref{Homogeneous-Section} we
present our results for the perturbation of homogeneous packings.
The perturbation of the packings that are settled under gravity is
investigated in Sec.~\ref{Gravity-Section}. The role of the
different preparation and perturbation methods are discussed in
Sec.~\ref{Discussions-Section}. Finally
Sec.~\ref{Conclusion-Section} concludes the paper.

\section{simulation method}
\label{SimulationMethod-Section}

We perform contact dynamics simulations on 2D granular packings of
cohesionless perfectly rigid disks. The numerical results of the
local perturbations are presented for two distinct settings: (i)
homogeneous random packings confined by an external pressure in
the absence of gravity [Fig.~\ref{Fig-SchematicPicture}(a)] and
(ii) packings confined by gravity
[Fig.~\ref{Fig-SchematicPicture}(b)]. In both cases, the numerical
experiments consist of two steps. First we prepare static
configurations of grains; then, we probe the packings by
perturbing their local structure. We apply the contact dynamics
method \cite{Jean99,Brendel04} for both procedures. The details of
the numerical methods are described for homogeneous and
inhomogeneous settings in
Secs.~\ref{SimulationMethod-Homogeneous-Subsection} and 
\ref{SimulationMethod-Gravity-Subsection}, respectively.

In the rest of the paper we have the following conventions. The
unit of the length is set to the maximum grain radius. As we
examine 2D systems, the disks have polydispersity to avoid
crystallization characteristic for two-dimensional monodisperse
ensembles. We use a uniform distribution of the disk radii over
the range between 0.5 and 1. The unit of the mass is set by
assuming that the material of the grains has unit density and the
masses of the disks are proportional to their areas.

\subsection{Homogeneous random packings}
\label{SimulationMethod-Homogeneous-Subsection}

We first examine the homogeneous configurations of disks. Here,
the acceleration of gravity is set to zero in order to avoid force
gradients in the samples. The number of the grains contained by
the packings ranges from $500$ to $8000$. As mentioned above, we
first generate static dense random packings by compressing the
initial configuration of the particles into a smaller space. The
compaction starts with a square box filled with loose granular
gas. The disks are initially placed at random without overlaps. We
apply periodic boundary conditions to avoid wall effects. In order
to achieve homogeneous packings, we generalized the method
proposed by Andersen \cite{Andersen80} to contact dynamics (for
details see \cite{Shaebani08jcp}). The main idea in this method is
that, instead of using pistons, compaction is achieved by imposing
a constant external pressure $P_\text{ext}$ and let the volume of
the cell evolve in time. In fact, the volume of the system, which
acts as an additional degree of freedom, couples with a confining
pressure bath, so that the volume change is controlled by the
difference between the external and internal pressures.

As the size of the cell shrinks due to the difference between
$P_\text{ext}$ and the internal pressure $P_\text{in}$, at some
point the grains cannot avoid touching each other anymore, and
start building up an inner pressure to avoid interpenetration.
Finally all motion stops because the grains block further
compaction. The sample is considered to have converged to
mechanical equilibrium when further time evolution leads to
negligible changes in the particle positions. At this point we
have a static jammed configuration under external pressure. The
mechanical equilibrium is achieved for each grain and the
corresponding $P_\text{in}$ equals $P_\text{ext}$. It is worth
noting that the packing configurations depend on the friction
coefficient $\mu$. We construct a new packing for each value of
$\mu$ before starting the perturbation process.

The perturbation is carried out in the following way: We choose
two adjacent particles in contact and force them to move apart
[Fig.~\ref{Fig-SchematicPicture}(a)]. As this case has been
described in \cite{Shaebani07}, here only a short review is given.
At the perturbation point we enforce the contacting surfaces to
open up to a small gap and determine the force that is needed to
fulfill this constraint (critical force). This concept is suited
very well to the contact dynamics method where interparticle
forces are handled as constraint forces, i.e., they are
calculated based on constraint conditions which prescribe the
relative motion of the contact surfaces \cite{Jean99,Brendel04}.
With enforcing the opening of the contact, we bring the system
immediately to the yield point where the perturbation induces
sliding and/or opening of some contacts and initiates collective
rearrangements of the particles at least in the vicinity of the
perturbation point. It is beneficial to choose small gap size as
we are interested in the onset of motion, how the static structure
breaks due to the perturbation. Large deformations, e.g., creation
of new contacts, are out of the scope of the present study. We
checked that for small gap sizes the displacement field (up to a
constant factor) and the critical force become independent of the
size of the gap. Our numerical measurements are performed in this
gap-independent region; the size of the gap $\xi$ is set to
$10^{-9}$ \cite{Shaebani07}. This value is far larger than the
displacement scale $10^{-14}$ that arises from the noise level of
particle velocities.

The perturbation process can be performed under two different
boundary conditions. One can either impose the fixed external
pressure condition in continuation of the assembling process, or
impose the fixed volume condition. In the former method, the
pressure of the system is constant and the volume is allowed to
change during the perturbation while in the latter method, the
volume is fixed and the pressure changes due to perturbation. This
paper mainly contains the results of perturbation method with
fixed pressure even though we compare some numerical results of
both methods in Sec.~\ref{FixedPressureVolumeMethods-Subsection}
and find no significant differences.

\subsection{Packings confined by gravity}
\label{SimulationMethod-Gravity-Subsection}

In the inhomogeneous case, the system is settled under gravity and
no pressure bath is used. Consequently there exists a pressure
gradient in the vertical direction. We present simulations on
model systems of $N=1600$ polydisperse disks. The system is
spatially periodic in the horizontal direction, and a 
one-dimensional chain of disks with random radii is fixed at the
bottom of the box to provide a rough bed [see
Fig.~\ref{Fig-SchematicPicture}(b)]. The particle-particle and
particle-base friction coefficients are the same. The starting
configuration is a dilute granular medium which consists of 
randomly distributed nonoverlapping particles with zero 
velocities. The average initial packing fraction $\phi$ ranges 
from $0.32$ to $0.40$ for different packings. Next the system is
allowed to settle on top of the rough bed under the influence of
the gravitational acceleration $g$. We wait until the packing
relaxes into equilibrium. The average static packing height ranges
from $35.2 \pm 0.1$ (approximately $23$  layers of grains) for 
samples with small friction coefficient $\mu = 10^{-8}$ to 
$36.9 \pm 0.1$ ($\sim 25$ layers) for samples with large friction 
coefficient $\mu = 10$. This construction method mimics the 
pouring of grains through a sieve far away from lateral walls.

To investigate the effect of friction and also to measure the
displacement response with better statistics, $20$ packings are
constructed and perturbed for each value of the friction
coefficient $\mu$.

After that we turn to the perturbation part where we perturb the
topmost or lowermost particles in two different experiments in
order to test the effect of local perturbation in the
inhomogeneous system [Fig.~\ref{Fig-SchematicPicture}(b)]. In the
first case, we choose a particle at the free surface of the
packing and force it to move vertically downwards by a small
displacement $\xi$. We measure the generated displacements of the
particle centers and also the force response of the system on the
perturbed particle. This latter is the vertical component of the
sum of the contact forces acting on the perturbed particle, which
plays the role of the critical force.

In the case where the system is perturbed from the bottom we
choose a particle at the lowermost part of the sample and force it
to move vertically upwards by a small upwards shift $\xi$. The
displacement pattern and the critical force are measured. In both
cases, the magnitude of the displacement $\xi$ is set to the same
value as the gap for the homogeneous system.

\section{Perturbation of homogeneous packings}
\label{Homogeneous-Section}

In this section, we will analyze the mechanical response of the
homogeneous packing to the local perturbation which we introduced
in Sec.~\ref{SimulationMethod-Homogeneous-Subsection} and see how
the response changes with the friction coefficient. The results
presented in this section belong to the system size $N=3000$
unless explicitly stated otherwise.

After opening up a contact that is selected for the perturbation,
we study the generated displacement field of particle centers and
the perturbation force which is needed to open up the contacting
surfaces at the perturbation point. This critical force $\Fcrit$
characterizes the strength of the system against the local
perturbation. Furthermore, the numerical results are compared for
the fixed pressure and the fixed volume perturbation methods. We
close this section by reporting the results of the generated force
and stress response fields.

\subsection{Displacement response field}
\label{DisplacementField-Subsection}

Our aim here is to find out how far the rearrangements have to
penetrate into the packing as a consequence of the prescribed
local deformation. Is there a related length scale?

\begin{figure}
\centerline{%
\epsfig{figure=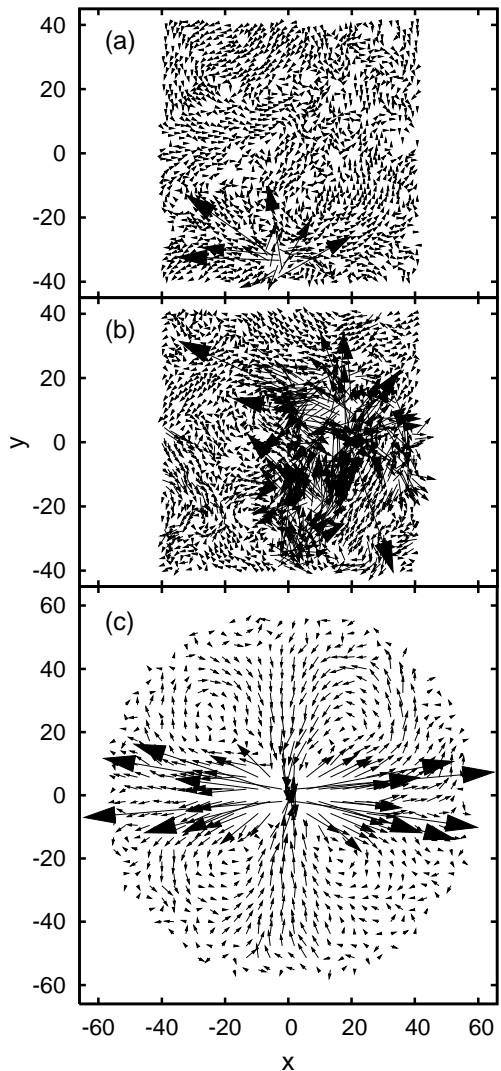,width=0.80\linewidth}}
\caption{(a),(b) Displacement response fields in the laboratory
frame for two different locations of the perturbation. The
resulting vector field can be relatively localized (a) or more
widespread (b) depending on the perturbed contact. (c)
Displacement response field in the contact frame averaged over
several thousand perturbations. The system contained $3000$ disks
with friction coefficient $0.5$. The unit of the length, in which 
$x$ and $y$ are measured, is set to the maximum grain radius. For 
clarity, the magnitude of the displacements has been increased by 
a factor of $10^{11}$ in all figures. 
\label{Fig-DisplacementField}}
\end{figure}

Our results reveal that the displacement of particle centers due
to the single contact perturbation form a disordered vector field.
This response field can be relatively localized
[Fig.~\ref{Fig-DisplacementField}(a)] or more widespread
[Fig.~\ref{Fig-DisplacementField}(b)] depending on the location of
perturbation.

As a measure for the magnitude of the displacement response, we
define the \emph{penetration depth} $\delta$ as

\begin{equation}
\delta = \frac{\displaystyle\sum_\text{i=1}^\text{N} |\vec
d_\text{i}| |\vec{r}_\text{i} \cdot \vec{n}|}
{\displaystyle\sum_\text{i=1}^\text{N} |\vec d_\text{i}|},
\label{delta-definition}
\end{equation}
where the sum runs over all particles, $\vec d_\text{i}$ is the
displacement vector of the $i$th particle center,
$\vec{r}_\text{i}$ is the distance vector from the perturbed
contact to the $i$th particle center, and $\vec n$ is the unit
contact normal of the perturbed contact.

The penetration depth $\delta$ characterizes the size of the
rearrangement zone in the direction of the contact normal.
$\delta$ exhibits large fluctuations depending on the perturbed
contact. For the displacement fields shown in
Figs.~\ref{Fig-DisplacementField}(a) and 
\ref{Fig-DisplacementField}(b) the values of $\delta$ are $7.68$ 
and $23.31$, respectively. The average penetration depth $\langle 
\delta \rangle$, which we calculated based on the perturbation of 
$1500$ randomly chosen contacts of the same system, is $19.8 \pm 
0.1$.

\begin{figure}
\epsfig{figure=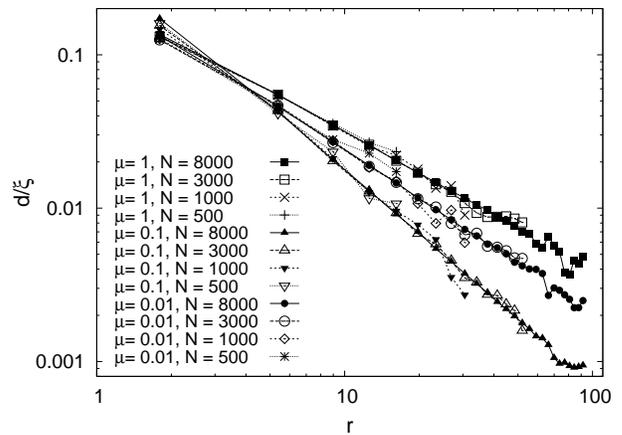,width=0.99\linewidth}
\caption{The magnitude of the average displacements $d$ in terms
of the distance from the perturbed contact $r$. Here $\xi$ stands
for the gap generated at the perturbed contact. Different slopes
correspond to different friction coefficients $\mu$. For each
value of friction four systems of different sizes are
investigated. The total number of particles is between $500$ and
$8000$. The error bars remain below $5 \%$ of the values 
$d(r) / \xi$ for the whole range of $r$. 
\label{Fig-DisplacementDistance}}
\end{figure}

In order to study the average properties of the displacement
fields, we perturb all contacts one by one, always starting with
the original static packing. In each case the particle movements
are recorded in the local contact frame where the perturbed
contact sits in the origin and the $x$ axis is chosen  parallel to
the contact normal, i.e., $x$ indicates the direction of the
separation, then we calculate the average displacement field.

Figure \ref{Fig-DisplacementField}(c) shows a smooth displacement
field obtained by averaging over all perturbed contacts. The
circular shape in Fig.~\ref{Fig-DisplacementField}(c) is achieved
because the original square shape of the system has many different
orientations when transformed into different contact frames.
Similar quadrupolar structures as in
Fig.~\ref{Fig-DisplacementField}(c) were found for shearing
amorphous systems, where localized quadrupolar rearrangements
appear at the onset of plastic events \cite{Maloney06}.

The magnitude of the average displacement vectors $d$ decays with
the distance $r$ from the perturbation point as shown in
Fig.~\ref{Fig-DisplacementField}(c). In order to investigate its
decay, we calculate $d$ by averaging out the angle of the
position. Figure \ref{Fig-DisplacementDistance} shows that the
magnitude of rearrangements $d$ decays as a power law of the
distance $r$,
\begin{equation}
  \label{powerlaw}
  d \propto  r^{-\alpha}.
\end{equation}

\begin{figure}
\centerline{%
\epsfig{figure=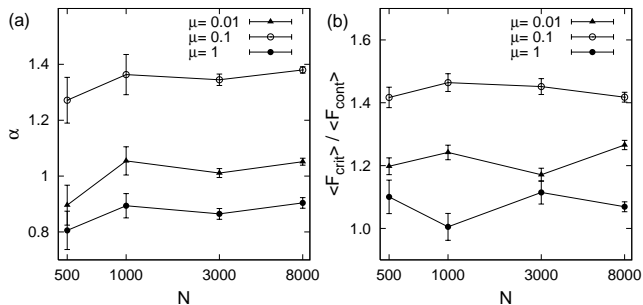,width=0.99\linewidth}}
\caption{Dependence of the penetration exponent $\alpha$ (a) and
the mean critical force $\langle \Fcrit \rangle$ (b) on the system
size $N$ for three different friction coefficients $\mu$.
\label{Fig-ForceAlphaSystemSize}}
\end{figure}

Different slopes in Fig.~\ref{Fig-DisplacementDistance} correspond
to different friction coefficients. However, the exponent $\alpha$
is approximately independent of the system size. This can be seen
more clearly in Fig.~\ref{Fig-ForceAlphaSystemSize}(a), in which
the exponent is shown in terms of the total number of particles $N$ 
for three different frictions. Our results show that $\alpha$
lies in the range $0.7-1.4$. The results of a recent
experiment performed by moving an intruder in a system of disks
\cite{Kolb06} displays similar power-law behavior with the same
range of $\alpha$.

The power-law decay of the average displacements indicates that
there is no characteristic size for the rearrangement zone;
instead, a decay exponent may be more suitable to characterize the
particle movements. Therefore the rearrangement region is not
bounded on sides by a penetration length and the quantity $\delta$
may not remain finite for an infinitely large system. Despite of
these facts, $\delta$ is still a useful measure of the
displacements for finite systems. Using $\delta$, one can still
compare two displacement fields for the same system size. Larger
$\delta$ means a larger rearrangement zone. Moreover, $\delta$ can
be easily measured also for single perturbations where the usage
of $\alpha$ would be troublesome. The exponent $\alpha$ works well
for the average displacement field but it is not a well defined
quantity for single perturbations where the fluctuation of $d(r)$
is so large that it cannot be fitted with a power-law decay.

To investigate the role of friction, we perturb several packings
constructed already with different friction coefficients, $\mu$.
The average penetration depth $\langle \delta \rangle$ has a
strong dependence on the friction coefficient $\mu$. It is a
nonmonotonic function with a sharp minimum at $\mu \approx 0.1$
[Fig.~\ref{Fig-FcritDeltaFriction-Homogeneous} (solid circles)].
Equivalent behavior is found for $-\alpha$ as a function of $\mu$
in Ref.~\cite{Shaebani07}. When the friction is increased starting
from zero, the decrease of $\langle \delta \rangle$ indicates that
the induced rearrangements become more localized. At $\mu \approx
0.1$ the process takes a sharp turn and further increase of the
friction leads to delocalization.

\subsection{Critical force}
\label{CriticalForce-Subsection}

We now turn our attention to the critical force $\Fcrit$. The
results show that $\Fcrit$ depends also strongly on the place of
the perturbation. First we check its average properties.

\begin{figure}
\epsfig{figure=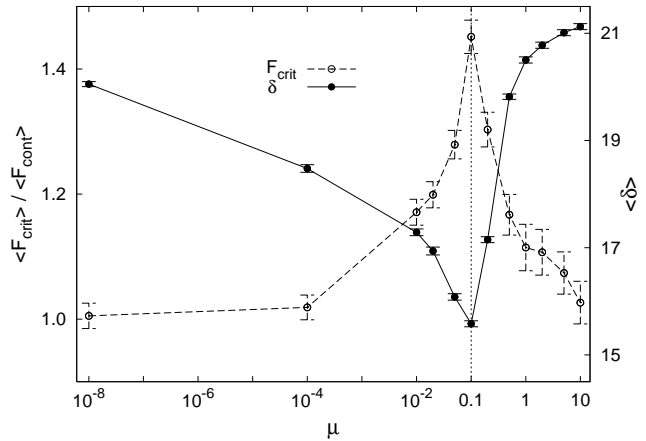,
width=0.99\linewidth} \caption{Average critical force $\langle
\Fcrit \rangle$ (open circles) scaled by the average normal
contact  force $\langle \Fcont \rangle$ and the average
penetration depth $\langle \delta \rangle$ (full circles) as
functions of the friction coefficient $\mu$. The vertical dashed
line emphasizes that the two extrema are located at the same
$\mu$. \label{Fig-FcritDeltaFriction-Homogeneous}}
\end{figure}

The average critical force $\langle \Fcrit \rangle$ again shows
strong dependence on the friction coefficient
[Fig.~\ref{Fig-FcritDeltaFriction-Homogeneous} (open circles)] and
remains approximately constant under changing the system size
[Fig.~\ref{Fig-ForceAlphaSystemSize}(b)]. Here, $\langle \Fcrit
\rangle$ is scaled by the average normal contact force $\langle
\Fcont \rangle$. In $\langle \Fcont \rangle$, only the normal
component of the contact forces is taken into account and the
average $\langle \cdot \cdot \cdot \rangle$ is taken over all 
the contacts of the given packing.

Both the critical force and the penetration depth show the
characteristic nonmonotonic behavior
(Fig.~\ref{Fig-FcritDeltaFriction-Homogeneous}) in contrast to
other quantities that describe the properties of the packing.
E.g., the average coordination number $z$, the average contact
force $\langle \Fcont \rangle$, and the packing fraction exhibit
smooth and monotonic functions of the friction coefficient with
plateaus for the low and high friction regions \cite{Shaebani07}.
$\Fcrit$ and $\delta$ behave completely differently. They exhibit
sharp extrema at the same friction: At $\mu \approx 0.1$ the
maximum critical force and the most localized rearrangement zone
is observed which has the meaning that the packing constructed
with friction $\mu=0.1$ is the most stable packing against local
perturbations. Moving towards the higher or lower friction
coefficients, packings get weaker against the perturbation, the
critical force becomes smaller, and the induced rearrangements
become more widespread.

\begin{figure}[b]
\centerline{%
\epsfig{figure=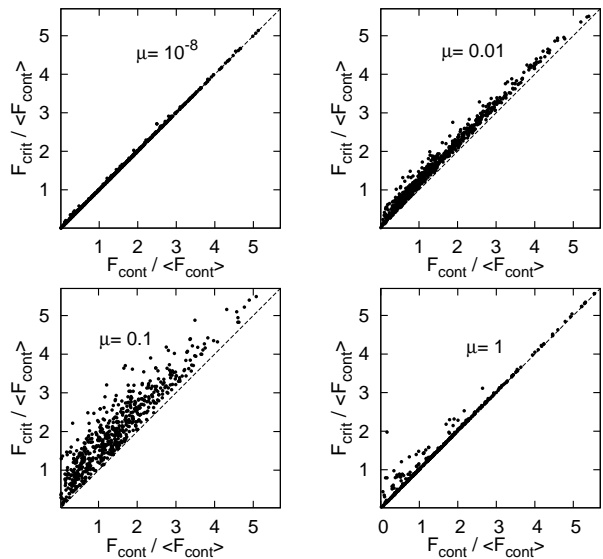,width=1.23\linewidth}}
\caption{Each data point represents one contact in the frame of
the critical force $\Fcrit$ and the normal component of the
original contact force $\Fcont$. The four figures correspond to
four packings constructed with different frictions. $\langle
\Fcont \rangle$ is the average normal contact force in each
packing. The dashed lines correspond to $\Fcrit = \Fcont$.
\label{Fig-FcritFcont}}
\end{figure}

\begin{figure}
\epsfig{figure=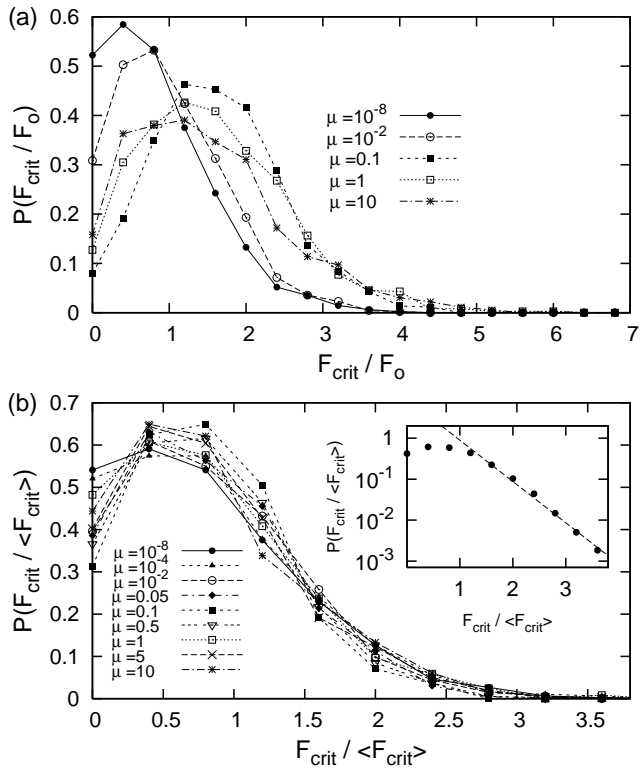,width=0.99\linewidth}
\caption{Probability distribution of the critical forces $\Fcrit$
(a) and the critical forces normalized with respect to their mean
in each packing $\Fcrit/\langle \Fcrit \rangle$ (b) for different
friction coefficients. The inset displays a semilogarithmic plot
of probability distribution of the normalized critical forces when
averaged over all friction coefficients. The dashed line is an
exponential fit of the tail of the curve.
\label{Fig-ForceDistributions}}
\end{figure}

To get more insight into the force response we examine $\Fcrit$ at
every contact separately. Figure \ref{Fig-FcritFcont} reveals that
the critical force is strongly correlated with the original
contact force. For small values of $\mu$ a pair of contacting
particles cannot resist a force of separation larger than the
force itself that originally presses the two contact surfaces
together. This can be understood well in the case of zero friction
where the structure is isostatic \cite{JNRoux00}, where the
structure and the external load $p_\text{ext}$ determine uniquely
what equilibrium force is acting between a selected pair of
contacting particles. If one pushes the two particles apart with a
larger force, it should be compensated by a negative contact
force. As the contact cannot exert negative forces we lose one
constraint and, consequently, one floppy mode \cite{JNRoux00}
appears in the system allowing collective motion of the particles.

Naturally, the critical force never falls below the actual contact
force, however, for the frictional case it can get larger. This can be
best seen for $\mu=0.1$ in Fig.~\ref{Fig-FcritFcont}.  For further
increase of $\mu$ the picture becomes similar to the case of zero
friction, i.e., the force response of a contact against opening is
basically given by the normal component of the original contact
force.

It is known that contact forces in random packings of frictional
rigid grains are not determined uniquely by the mechanical
equilibrium and Coulomb condition
\cite{Snoijer04,Unger05,McNamara05,Ostojic06}. There is an
ensemble of admissible force networks that satisfy all of these
conditions in the same configuration of grains. The extent of the
force indeterminacy as a function of the friction coefficient
$\mu$ was numerically examined in \cite{Unger05} for packings of
rigid disks, where a nonmonotonic friction dependence was found
with a maximum value at $\mu \approx 0.1$. A direct connection
between the critical force $\Fcrit$ and the extent of the
indeterminacy at a given contact is established in
\cite{Shaebani07} where it was found that the critical force
equals the maximum possible contact force at the same contact
taken over the ensemble of admissible force networks. For further
details on the consequence of force indeterminacy the reader is
referred to \cite{Shaebani07,Unger05,Shaebani08pre2}.

\begin{figure}
\epsfig{figure=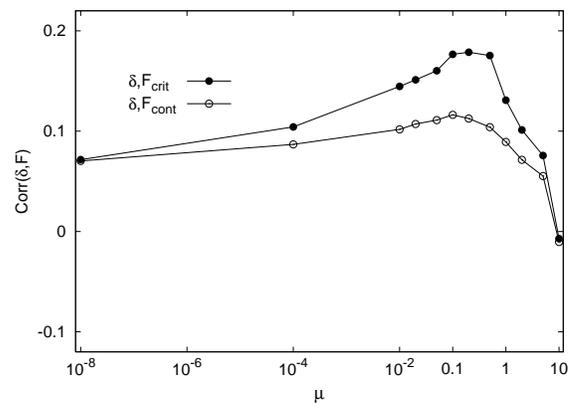,
width=0.9\linewidth} \caption{The correlation between the critical
force $\Fcrit$ (full circles) or the normal component of the
original contact force $\Fcont$ (open circles) and the penetration
depth $\delta$ in terms of the friction coefficient $\mu$.
\label{Fig-Correlation-Delta-F}}
\end{figure}

As we mentioned before, the value of the critical force widely
changes depending on the perturbed contact. The probability
distributions for $\Fcrit$ are displayed in
Fig.~\ref{Fig-ForceDistributions}(a) for different friction
coefficients. The critical forces are scaled in units of
$F_\text{0}$ set by the external pressure and by the average
radius of the particles, $F_\text{0} = 2 R_\text{avg}
P_\text{ext}$. Here, we use the unit $F_\text{0}$ because, unlike
$\langle \Fcont \rangle$, it provides a fixed force scale for all
systems which is independent of the friction coefficient.
Figure \ref{Fig-ForceDistributions}(a) shows that the probability
distributions depend strongly on $\mu$. With increasing friction,
probability distributions become broader and the curves are
shifted. The shift is nonmonotonic as expected from the behavior
of $\langle \Fcrit \rangle$. The curves move rightwards and then
leftwards below and above $\mu = 0.1$, respectively.

\begin{figure}
\epsfig{figure=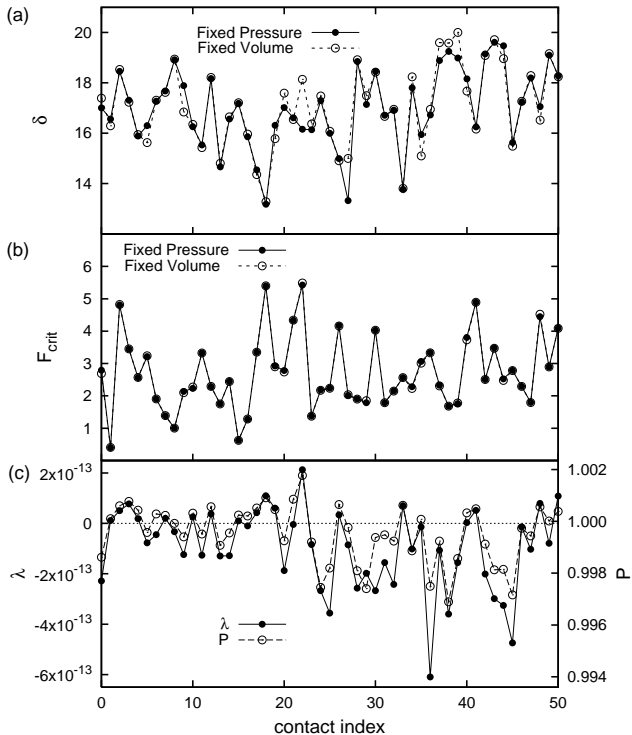, width=0.99\linewidth}
\caption{Comparison of the results of the fixed pressure (solid
circles) and the fixed volume (open circles) perturbation methods.
(a) The penetration depth $\delta$ and (b) the critical force
$\Fcrit$ for several contacts subjected to the perturbation. (c)
The relative change in the size of the system $\lambda$ (solid
circles) and the pressure $P$ (open circles) obtained during the
perturbation of each contact with the fixed pressure and fixed
volume methods, respectively. Here, the friction coefficient is
$0.5$. \label{Fig-FixedVolumePressure}}
\end{figure}

When normalized with respect to their mean values, these
distributions are approximately independent of the friction
coefficient [Fig.~\ref{Fig-ForceDistributions}(b)] and their tail
can be fitted with an exponential decay

\begin{equation}
P( \Fcrit/\langle \Fcrit \rangle )\propto e^{(-\beta) \Fcrit/\langle 
\Fcrit \rangle}
\end{equation}
where $\beta = 2.3\pm0.1$ when averaged over all friction
coefficients [see the inset of
Fig.~\ref{Fig-ForceDistributions}(b)].

The tail of the curves in Fig.~\ref{Fig-ForceDistributions}(b) for
the critical forces is reminiscent of the tail of the probability
distributions of the contact forces which has been studied
extensively in the literature
\cite{Mueth98,Radjai96,Goldenberg03,Majmudar05,vanEerd07}. This
similarity can be understood well, based on the strong correlation
between the critical force and the original contact force which is
described in Fig.~\ref{Fig-FcritFcont}.

Next we focus on the correlation between the critical force
$\Fcrit$ and the penetration depth $\delta$.
Figure \ref{Fig-Correlation-Delta-F} (solid circles) displays that
there is a weak correlation between these two quantities with a
maximum value again around $\mu=0.1$ and it vanishes for large
friction coefficients. The existence of the correlation means
that, on average, a slightly larger rearrangement zone is expected
for a larger critical force. Open circles in
Fig.~\ref{Fig-Correlation-Delta-F} reveal smaller correlations
between the normal component of the original contact force
$\Fcont$ and $\delta$.

\subsection{Fixed pressure and fixed volume perturbations}
\label{FixedPressureVolumeMethods-Subsection}

In Sec.~\ref{SimulationMethod-Homogeneous-Subsection} we mentioned
that one can perturb the system by imposing either the fixed
external pressure condition or the fixed volume condition. In this
section we compare these two perturbation methods by applying them
in the same test system on the same list of contacts. The
penetration depth $\delta$ and the critical force $\Fcrit$ are
shown for each perturbed contact in
Figs.~\ref{Fig-FixedVolumePressure}(a) and 
\ref{Fig-FixedVolumePressure}(b), respectively. The results of the 
two methods are very similar to each other even though the imposed 
boundary conditions are basically different.

In the fixed volume method the pressure $P$ is allowed to change, 
while in the fixed pressure method the variable quantity is the
volume of the system. Let us measure the volume change by $\lambda
= \Delta L / L$, where $L$ is the size of the system. Figure 
\ref{Fig-FixedVolumePressure}(c) shows that there is a strong
correlation between the variable quantities $P$ for one method and
$\lambda$ for the other. The expansion (contraction) of the system
due to the perturbation of a contact with the fixed pressure
method corresponds to pressure increase (decrease) when perturbing
the same contact with the fixed volume method.

\subsection{Force and stress response fields}
\label{ForceStressResponse-Subsection}

\begin{figure}
\epsfig{figure=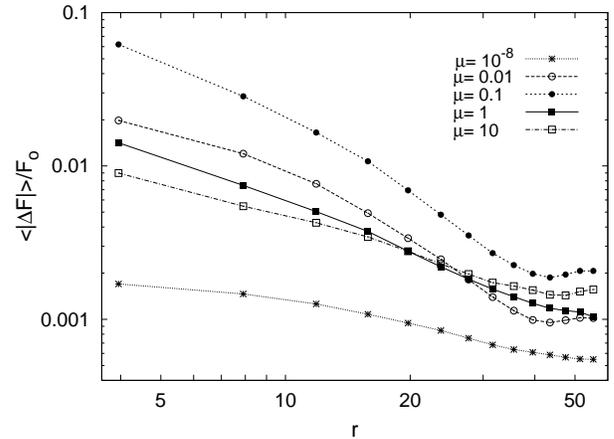, width=0.99\linewidth} 
\caption{The average magnitude of the change
in the normal component of contact forces due to the perturbation
$\langle |\Delta F| \rangle$ scaled by $F_\text{0}$ in terms of
the distance from the perturbed contact $r$. Different curves
correspond to different friction coefficients $\mu$.
\label{Fig-ForceResponse-Distance}}
\end{figure}

Finally, we investigate how other contact forces and the
corresponding stress field is changed by the local perturbation.
We measure the average magnitude of the change in the normal
component of the contact forces $\langle |\Delta F| \rangle$ which
is caused by the perturbation.
Figure \ref{Fig-ForceResponse-Distance} shows $\langle |\Delta F|
\rangle$ as a function of the distance from the perturbed contact.
These curves, unlike the displacement response, do not follow a
power law decay. Close to the perturbation point, much larger
$\langle |\Delta F| \rangle$ is observed for $\mu=0.1$ than for
the extreme values of friction ($\mu=10^\text{-8}$, $10$).
Depending on the friction, $\langle |\Delta F| \rangle$ can be
changed even by a factor $30$ in the vicinity of the perturbation
point. The values of $\langle |\Delta F| \rangle$ become closer to
each other for different frictions far away from the perturbation
point since the decay of $\langle |\Delta F| \rangle$ becomes
steeper for intermediate values of friction. Apparently $|\Delta
F|$ goes to zero with $r$ leading to tiny changes in the contact
forces far away from the perturbation point. Interestingly, even
these tiny forces are able to break the solid state of the packing
and induce rearrangements of the particles
(Fig.~\ref{Fig-DisplacementField}).

We calculate the average stress field caused by the perturbation,
measured always in the contact frame and averaged over several
thousands of perturbations. We divide the system into square grid 
boxes of size $2$. This corresponds to the maximum diameter or twice 
the minimum diameter of the particles. The whole stress we achieved 
for each box is assigned to the position of the center of the box. 
The stress tensor $\sigma_{\alpha \beta}$ in each grid box is given 
\cite{Christoffersen81} by
\begin{equation}
\sigma_{\alpha \beta} = \frac{1}{V} \displaystyle\sum_{i<j}
\theta_\text{ij} f^{\alpha}_\text{ij} r^{\beta}_\text{ij},
\end{equation}
where $V$ is the area of the box, $f^{\alpha}_\text{ij}$ is the 
$\alpha$ component of the force exerted on particle $i$ by 
particle $j$, and the vector $\vec r_\text{ij}$ points from the 
center of particle $i$ to the center of $j$ (one has to take the 
periodic boundary conditions into account and involve nearest 
image neighbors). The sum runs over all pairs of contacting 
particles. $\theta_\text{ij}$ is a number between $0$ and $1$ that 
gives what fraction of the line segment connecting the centers of 
the two particles is located in the box. Thus if the line segment 
is cut by grid lines the stress contribution of the contact is 
divided among the corresponding boxes.

First we investigate the pressure which is defined as half of the
trace of the stress tensor. Originally (before perturbation) the
pressure is spatially constant due to the symmetry of the
compression process. To investigate the effect of perturbation on
the local pressure, we calculate the pressure change as
\begin{equation}
\Delta P = \frac{1}{2} \text{tr} (\sigma-\sigma_\text{0}),
\end{equation}
where $\sigma_\text{0}$ is the stress tensor before the
perturbation.

\begin{figure}
\centerline{%
\epsfig{figure=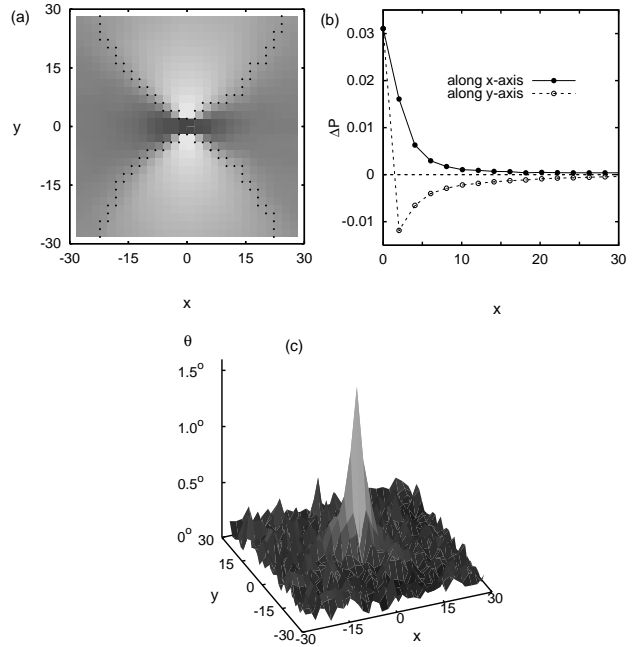,width=0.99\linewidth}}
\caption{(a) The profile of the pressure change $\Delta P$.
Shading is logarithmic in the amplitude of the pressure change,
with black indicating a pressure increase, and white, a pressure
decrease. Dots show where the sign of $\Delta P$ is changed. (b)
$\Delta P$ along the $x$ and $y$ axes. (c) The maximum angle of
friction mobilization $\alpha_\text{max}$. The $x$ axis indicates
the direction of separation at the perturbed contact. The friction
coefficient is $0.5$. \label{Fig-Pressure}}
\end{figure}

Figure \ref{Fig-Pressure}(a) shows a logarithmic shading of the
pressure change with black indicating a pressure increase, and
white, a pressure decrease. This figure reveals that the pressure
change decays fast with the distance from the perturbation point.
The borders between the regions of positive and negative $\Delta
P$ are also indicated in Fig.~\ref{Fig-Pressure}(a). $\Delta P$ is
positive (negative) along (perpendicular to) the perturbation
direction. For ease of comparison Fig.~\ref{Fig-Pressure}(b) shows
$\Delta P$ along and perpendicular to the $x$ axis in the contact
frame.

Next we investigate the angle of shearing, which is usually used
to describe how close the material is to plastic deformation. The
angle of shearing $\theta$ is provided by the local stress tensor
\cite{Nedderman92}:
\begin{equation}
\theta= \arcsin \left(\frac{\tau_\text{max}}{P} \right)
                =\arcsin \left(\frac{\sqrt{\sigma^\text{2}_\text{xy} +
                (\sigma_\text{xx}-P)^\text{2}}}{P}\right).
\end{equation}
When the shear stress is measured in an imaginary plane at some
point of the material, its value depends on the orientation of
the plane. For a given stress state, $\tau_\text{max}$ denotes
the maximum shear stress among all orientations. Dry granular
media are often characterized by a critical angle of shearing
resistance $\theta_\text{crit}$ \cite{Nedderman92}. The material
is expected to behave as solid until the angle of shearing
remains below the critical angle and plastic deformation occurs
when the critical angle is reached.

Since the local stress is symmetric before the perturbation, the
initial value of the local angle of shearing $\theta_\text{0}$ is
approximately zero. $\theta$ induced by the perturbation exhibits
also a fast decay away from the perturbed contact.
Figure \ref{Fig-Pressure}(c) shows that $\theta$ is very small
throughout the system. Interestingly, even the largest value
$\theta \approx 1.5^\circ$ is far below the critical angle (the
typical value of $\theta_\text{crit}$ is around $20^\circ-30^\circ$),
still the perturbation is able to break the solid structure of the
packing.

\section{Perturbation of packings confined by gravity}
\label{Gravity-Section}

In this section we present the numerical results of the local
perturbations of packings confined by gravity. It it important to
note that both preparation and perturbation steps are performed in
the presence of gravity. In
Sec.~\ref{SimulationMethod-Gravity-Subsection} we explained how
such static configurations are prepared. We also described how we
perturb the topmost and lowermost particles in two different
measurements. Actually, these systems are more realistic in the
preparation and perturbation methods than those investigated in
Sec.~\ref{Homogeneous-Section}.

Our main aim is to study whether the nonmonotonic friction dependence 
of the mechanical response found for the ideal homogeneous 
packings is reproduced in these realistic systems. Here, by 
shifting a particle downwards at the free surface or
upwards at the bottom of the packing, we study the generated
rearrangements of the particle centers and also the critical force
on the perturbed particle.

We start our investigation with the results of the displacement
response field. We find that particle movements are not bounded to
a small vicinity of the perturbation point but we observe
displacements even in regions of the system that are far from the
perturbed particle. This indicates a long range effect similar to
those found for homogeneous packings
(Sec.~\ref{Homogeneous-Section}) and in experiments
\cite{Kolb04,Kolb06}. 

Here again, we characterize the size of the rearrangement zone in
our finite systems with the penetration depth $\delta$. Similarly
to Sec.~\ref{Homogeneous-Section}, $\delta$ is defined by the same
expression [Eq.~(\ref{delta-definition})] also in the present case,
only $\vec r_i$ now denotes the distance vector from the center of
the perturbed particle to the center of the $i$th particle and
$\vec n$ is pointing vertically downwards (upwards) for
top perturbations (bottom perturbations). Thus $\delta$ has the
meaning of the vertical size of the rearrangement zone. Similar to
the homogeneous case, $\delta$ depends strongly on the perturbed
particle; therefore we repeat the perturbation for many particles
to obtain the average value $\langle \delta \rangle$.

$\langle \delta \rangle$ is recorded separately for top and
bottom perturbations and for each value of friction. In each case,
the average value $\langle \delta \rangle$ is calculated over
approximately $1000$ perturbed particles. These particles are
selected as follows. We divide the width of the system into
several bins of size roughly equal to the average diameter of the
particles. Among the particle centers located at the same bin, we
find the highest and lowest centers and the corresponding two
particles are selected for the top and bottom perturbations,
respectively. We repeat this procedure for each bin.

It has to be noted that not all the selected particles for
bottom perturbation are taken into account in the calculation of
$\langle \delta \rangle$. We exclude rattler particles that do not
take part in the force transmission because they are screened by a
local arch. Of course, there are no true rattlers in the presence
of gravity, because every particle inevitably has force carrying
contacts due to its own weight. In the case of gravity we regard a
particle as a rattler if its upward perturbation generates no
force on the particle, i.e.\ if its critical force is zero.

\begin{figure}
\epsfig{figure=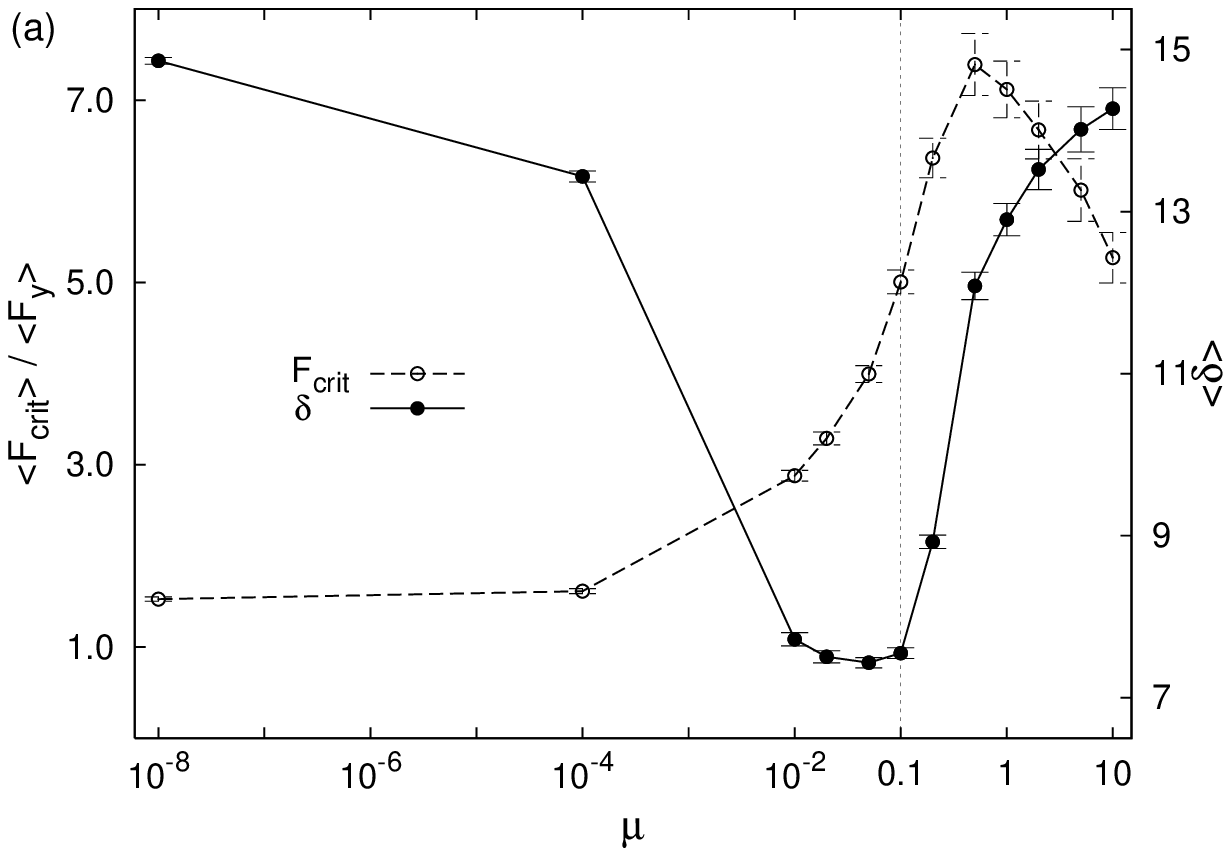, width=0.9\linewidth}
\epsfig{figure=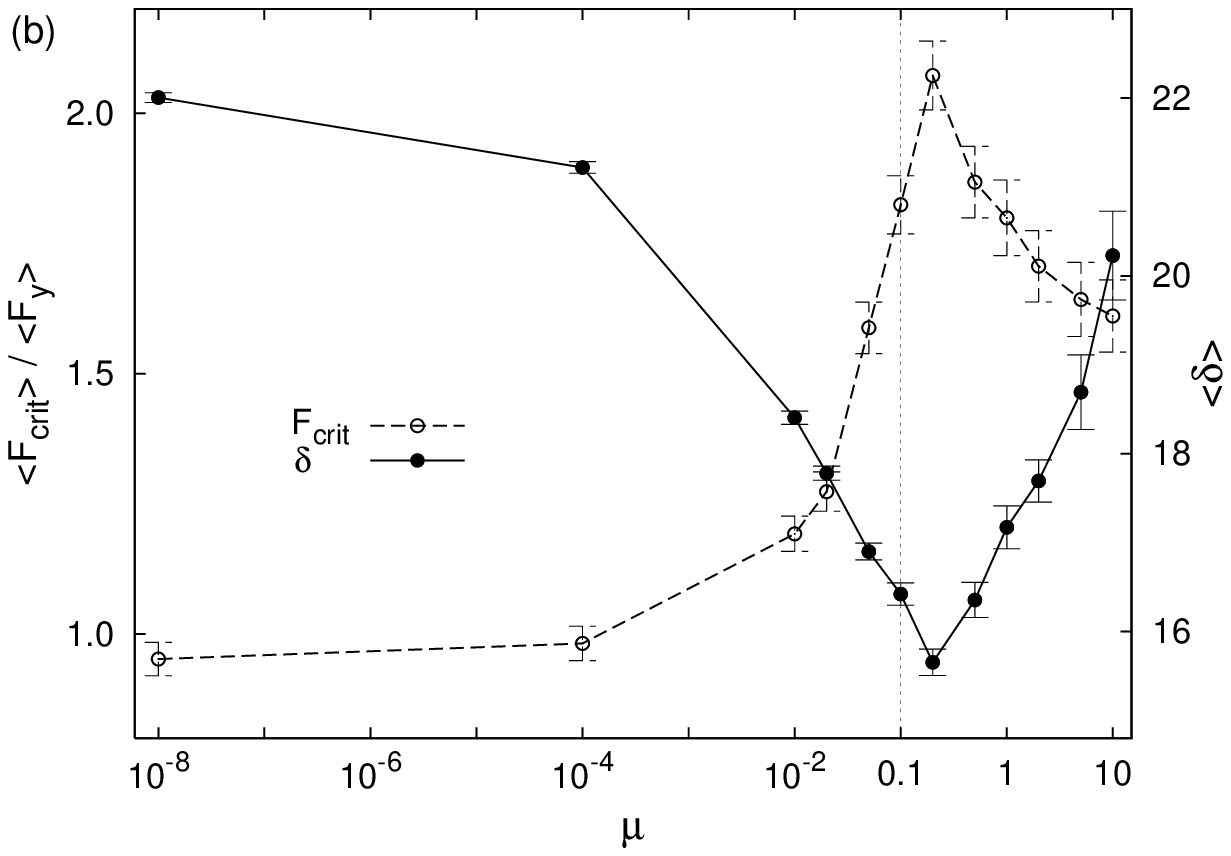, width=0.89\linewidth} 
\caption{The average penetration depth
$\langle \delta \rangle$ (solid circles), and the average critical
force $\langle \Fcrit \rangle$ (open circles) with respect to the
average effective weight $\langle F_y \rangle$ as functions of the
friction coefficient $\mu$, for the perturbation of topmost (a)
and lowermost (b) particles. The vertical dashed lines are
reminiscent of the position of the extrema in the homogeneous
case. \label{Fig-FcritDeltaFriction-Gravity}}
\end{figure}

To investigate the friction dependence of $\langle \delta \rangle$
we perturb several packings constructed with different friction
coefficients $\mu$.
Figure \ref{Fig-FcritDeltaFriction-Gravity} (solid circles) shows the
average penetration depth as a function of $\mu$ for both top and
bottom perturbations. It turns out that the qualitative behavior
in both measurements is similar to the homogeneous case:
nonmonotonic friction dependence with a minimum at intermediate
friction is found. However, the places of the minima are shifted
compared to the homogeneous case and also compared to each other.
The possible explanations will be discussed in the next section.

In fact, the bottom perturbation with gravity resembles much more
the homogeneous case than the top perturbation. For penetration
depths in Figs.~\ref{Fig-FcritDeltaFriction-Homogeneous}
and~\ref{Fig-FcritDeltaFriction-Gravity}(b) the minima are quite
sharp and the actual values of $\delta$ are almost the same
concerning the minimum $\delta$ and the values for small and large
$\mu$. For the top perturbation the situation is different. The
penetration depth is much smaller over the whole range of
friction, $\delta$ has a much broader minimum, and the ratio
between the maximum and minimum values ($\approx 2$) is much
larger than in the other two cases.

For the investigation of the critical force $\Fcrit$ we first have
to find a proper unit in which the average critical force is
measured in order to make the results comparable. This is needed
because originally the perturbed particles experience different
average load depending on whether they are located in the top or
in the bottom layer and also on magnitude of the friction. If the
original load on a particle is larger then the critical force is
expected to be larger as well.

We will use the quantity $\langle F_\text{y} \rangle$ as the force
unit, where $F_\text{y}$ is defined for a single particle as
follows:
\begin{equation}
F_\text{y} = \displaystyle\sum_{\lbrace \text{c} \vert
F_\text{y}^\text{c} > \text{0} \rbrace } F_\text{y}^\text{c}.
\end{equation}
Here $F_\text{y}^\text{c}$ is the $y$ (vertical) component
of the contact force exerted on the particle at contact
$c$, and the sum runs over contacts of the particle with
positive $F_\text{y}^\text{c}$, i.e., over the supporting contacts
that carry the particle against gravity. Similarly to $\langle
\delta \rangle$, the average $\langle \cdot \cdot \cdot \rangle$
here is taken over the perturbed particles and rattlers again are
excluded for bottom perturbations.

$F_\text{y}$ has the meaning of an effective weight of the
particle that is loaded on the supporting contacts below. It is
constituted by the own weight of the particle plus the weight of
other particles that is transmitted from above. Here we deal only
with vertical components of the contact forces because the
perturbation is performed in the vertical direction. This is in
analogy with Sec.~\ref{Homogeneous-Section} where the direction of
the perturbation was parallel to the contact normal therefore we
used the normal component of the original contact forces as the 
force unit.

The average effective weight $\langle F_\text{y} \rangle$ is
clearly different for the top and bottom perturbation due to the
pressure gradient. $\langle F_\text{y} \rangle$ depends also on
friction; it is an increasing function of $\mu$ for both cases.
For the lowermost particles, $\langle F_\text{y} \rangle$ ranges
from $28.4 \pm 0.9$ for packings with small friction coefficient 
$\mu = 10^{-8}$ to $40 \pm 1$ for packings with large 
friction coefficient $\mu = 10$. Corresponding values for the 
topmost particles are $1.25 \pm 0.03$ and $1.44 \pm 0.04$.

We record the critical force for each perturbed particle, where
the values of $\Fcrit$ show large fluctuations. We determine the
average $\langle \Fcrit \rangle$ separately for top and
bottom perturbations (rattlers are not taken into account in the
latter case). The average critical force is displayed in
Fig.~\ref{Fig-FcritDeltaFriction-Gravity} (open circles) scaled by
the average effective weight $\langle F_\text{y} \rangle$, where
the data show the dependence on friction for both types of
perturbation. We find that the nonmonotonic behavior with quite
sharp maximum, which was observed in
Sec.~\ref{Homogeneous-Section}, is reproduced here.

It is a common feature of
Figs.~\ref{Fig-FcritDeltaFriction-Gravity}(a) and 
\ref{Fig-FcritDeltaFriction-Gravity}(b) that $\langle
\Fcrit \rangle$ is close to the average effective weight $\langle
F_\text{y} \rangle$ in the small friction limit and the decreasing
branches of the curves at large frictions do not reach the same
level but, in contrast to the homogeneous case, $\langle \Fcrit
\rangle$ remains considerably larger. Another difference compared
to Fig.~\ref{Fig-FcritDeltaFriction-Homogeneous} is that the
maxima of $\langle \Fcrit \rangle$ are shifted rightwards.

Figure \ref{Fig-FcritDeltaFriction-Gravity} reveals also some
differences between the two types of perturbation we applied in
the presence of gravity. The scaled critical force, e.g., is much
larger in Fig.~\ref{Fig-FcritDeltaFriction-Gravity}(a) than in
Fig.~\ref{Fig-FcritDeltaFriction-Gravity}(b). Interestingly, the
variation range of the scaled critical force for the
bottom perturbation is very similar to the homogeneous case. It is
also shown in Fig.~\ref{Fig-FcritDeltaFriction-Gravity} that the
extrema of $\langle \Fcrit \rangle$ and $\langle \delta \rangle$
are located at different values of friction for top perturbation,
while for bottom perturbation the extrema are aligned similarly to
Fig.~\ref{Fig-FcritDeltaFriction-Homogeneous}.

\section{Discussion}
\label{Discussions-Section}

In the previous two sections we applied localized perturbations in
a few different ways; we separated contacting particles and
shifted single particles at the free surface or at the bottom of
the system. We tested the critical force and the penetration depth
of the perturbations both in the presence and in the absence of
gravity. The observed behavior of these parameters was basically
the same in all cases: They show an interesting nonmonotonic
dependence on the coefficient of friction (see
Figs.~\ref{Fig-FcritDeltaFriction-Homogeneous}
and~\ref{Fig-FcritDeltaFriction-Gravity}). Of course, there are
also some differences between the curves presented in
Figs.~\ref{Fig-FcritDeltaFriction-Homogeneous}
and~\ref{Fig-FcritDeltaFriction-Gravity}. These differences may
have various origins; here we discuss some possible causes.

\subsection{Connectivity}

\begin{figure}
\epsfig{figure=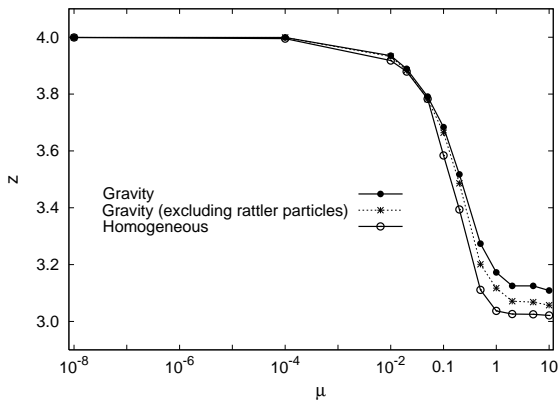,
width=0.9\linewidth} \caption{Influence of the friction on the
average coordination number $z$ for homogeneously confined
packings (open circles) and packings settled under gravity (full
circles). The middle curve (stars) corresponds to the gravity
case, when the contribution of the rattler particles, that
transmit no load except their own weight, is subtracted.
\label{Fig-CoordinationFriction}}
\end{figure}

We have used two different methods of preparation: ``homogeneous
compaction'' in which the grains are compressed by a confining
external pressure, and ``compaction by gravity'' where the grains
are piled due to gravitational acceleration. These different
preparation methods lead to different connectivity of the
packings. To verify this, we determine the average coordination
number $z=2N_\text{c}/N^\prime$, where $N_\text{c}$ is the total
number of contacts and $N^\prime$ is the number of particles that
have force carrying contacts. It can be seen in
Fig.~\ref{Fig-CoordinationFriction} that the preparation with
gravity provides a larger coordination number than the homogeneous
compaction. The deviation is significant for large friction
coefficients. In the region $\mu > 1$, $z$ approximately equals the
critical value $3$ for homogeneous compaction (open circles in
Fig.~\ref{Fig-CoordinationFriction}). This reveals that the
structure of the packing is very close to isostatic
\cite{Moukarzel98,JNRoux00,Unger05} where the contact forces are
uniquely determined by the equations of mechanical equilibrium of
the particles. The packings constructed by gravity are far from
being isostatic in the frictional case (full circles in
Fig.~\ref{Fig-CoordinationFriction}) therefore large indeterminacy
of the forces is expected \cite{Unger05}.

Larger force indeterminacy makes the packings more stable against
local perturbations and leads to larger critical forces
\cite{Shaebani07}. This explains why the critical force is
considerably larger for the right than the left side for both
Figs.~\ref{Fig-FcritDeltaFriction-Gravity}(a) and 
\ref{Fig-FcritDeltaFriction-Gravity}(b), while for
the homogeneous case approximately the same critical forces are
found in the small and large friction limits. This effect may also
cause the rightward shift of the maxima of $\langle \Fcrit
\rangle$ for the top and bottom perturbations in the presence of
gravity.

We note that the definition of $z$ applied here excludes rattlers
in zero gravity but in the presence of gravity all the particles
are taken into account, even those that effectively behave as
rattlers. It is important to point out that the difference in the
connectivity we found here cannot be traced back entirely to the
handling of rattlers. Even if we exclude rattlers (particles with
zero critical force in the upward perturbation) in the calculation
of $z$ we cannot achieve the isostatic limit $3$ for packings
settled under gravity. This is shown in
Fig.~\ref{Fig-CoordinationFriction} (stars) where the average
coordination number is determined only for the force carrying
structure, without rattlers.

The above discussed difference in the connectivity and its
consequences are observed only for the frictional particles. In
the zero friction limit we obtain the same coordination number
$z=4$ for both perturbation methods. This is the critical
coordination number for frictionless disks showing that inner
structure of these packings is isostatic \cite{Moukarzel98}.

\subsection{Pressure gradient}

The presence of pressure gradient in the case of gravity
(Sec.~\ref{Gravity-Section}) makes an important difference
compared to the homogeneous pressure
(Sec.~\ref{Homogeneous-Section}). The perturbation causes
displacements of the particles in a relatively large region. The
different parts of this rearrangement zone can experience
different local pressure. This effect is significant especially
for \emph{top perturbations} where the pressure grows
proportionally with vertical distance from the place of the
perturbation. The large relative change in the local pressure
suppresses rearrangements in deeper layers which leads to
considerably \emph{smaller penetration depths} and also to
\emph{larger critical forces} with respect to the related force
scale of the perturbed particles.

One can see this overall shift of $\langle \delta \rangle$ and
$\langle \Fcrit \rangle$ for top perturbation in the entire region
of $\mu$ [Fig.~\ref{Fig-FcritDeltaFriction-Gravity}(a)] when
compared to the homogeneous case
(Fig.~\ref{Fig-FcritDeltaFriction-Homogeneous}). For
bottom perturbations this effect is less important as the relative
change in the local pressure remains small in the vicinity of the
perturbed particle. This explains the smaller differences in
$\langle \delta \rangle$ and $\langle \Fcrit \rangle$ between
Figs.~\ref{Fig-FcritDeltaFriction-Homogeneous} and 
\ref{Fig-FcritDeltaFriction-Gravity}(b).

\subsection{Connection to force indeterminacy}

As mentioned in Sec.~\ref{CriticalForce-Subsection}, contact forces
are not determined statically in packings of frictional disks.
There exists an ensemble of force networks that solve the original
problem, i.e., they provide mechanical equilibrium under the given
external load and satisfy the Coulomb condition at every contact.
We can refer to this ensemble as the original solution set. Due to
the contact perturbation we applied in
Sec.~\ref{Homogeneous-Section} the packing finally chooses the
force network which contains the maximum possible contact force at
the perturbed contact \cite{Shaebani07}. In other words, the
perturbation drives the system into the border of the original
solution set. Therefore the critical force is directly connected
to the extent of indeterminacy of the contact forces.

It is important to point out that the relation to the original
force ensemble is different in the case of Sec.~\ref{Gravity-Section},
where we perturbed particles instead of contacts. Here the force
network generated by the perturbation corresponds to a different
external load and, consequently, it is located outside of the
original solution set. Therefore the critical force is not related
directly to the indeterminacy of the contact forces in case of top
and bottom perturbations. This seems to be a significant
difference between contact and particle perturbations which could
be one reason for the deviation observed for $\langle \Fcrit
\rangle$ between Secs.~\ref{Homogeneous-Section}
and \ref{Gravity-Section}.

\subsection{Other effects}

The systems constructed with homogeneous compaction are close to
isotropic \cite{Shaebani08jcp} in contrast to the packings settled
under gravity, where local stress and fabric are anisotropic due
to the special (vertical) direction of the compaction. In the
presence of gravity we perturbed the packings in this special
direction which may lead to different response properties compared
to the homogeneous case.

Moreover, for the vertical perturbations the up and down
directions may behave differently. The direction of gravity has
even a structural signature in the packing. E.g., the number of
contacts of a particle that have a vertical position higher than
the particle center exhibits larger fluctuations than the number
of contacts that are located below the center. One can mention
also the particles that are acting as rattlers when perturbed
upwards (zero critical force) but there is no rattler behavior for
downward perturbations where the critical force is always
positive. Therefore perturbations in the up and down directions
may lead to a different response even if the perturbations are 
performed for the same configuration of particles inside the bulk.

Another aspect that makes a difference between top and bottom
perturbations is the boundary condition. The top of the system is
not bounded, thus the displacement field generated by the
perturbation is allowed to pass through the original boundary.
This is not possible for the bottom perturbation, where the motion
of the surrounding particles is bounded by a rigid rough bed at
the bottom.

\section{Conclusion}
\label{Conclusion-Section}

In this work we presented the numerical results of the measurement
of mechanical response to localized perturbations. Based on
contact dynamics simulations we prepared 2D static granular
assemblies with and without gravity, then we perturbed the systems
with different methods to achieve the unjamming transition.
Despite all the effects that are expected to influence the
response of the packings depending on the preparation and
perturbation methods, the qualitative behavior seems to be very
robust. We found that both the resistance of the packings against
the perturbations and the penetration depth of the generated
displacement field are sensitive to the interparticle friction
coefficient: the surprising nonmonotonic dependence on friction is
reproduced in all cases that were studied in the present work.

The nonmonotonic behavior of the critical force in the case of 
contact perturbation can be understood based on the nonmonotonic 
indeterminacy of contact forces. However, the indeterminacy of 
forces seems to influence the critical force and the penetration 
depth generally. Further studies are needed to clarify this 
relationship.

\begin{acknowledgments}
We acknowledge support by grants No.~OTKA T049403, No.~OTKA
PD073172, and by the Bolyai Janos program of the Hungarian 
Academy of Sciences.
\end{acknowledgments}

\bibliography{Ref}

\end{document}